\documentclass[aps,pre,twocolumn,superscriptaddress,reprint]{revtex4-1}
\usepackage{graphicx, color}
\usepackage{amsmath}
\usepackage{amssymb}
\usepackage{amsfonts}
\usepackage{mathrsfs}
\usepackage{times}
\def\beq{\begin{equation}}
\def\eeq{\end{equation}}
\def\bea{\begin{eqnarray}}
\def\eea{\end{eqnarray}}
\def\be{\begin{equation}}
\def\ee{\end{equation}}

\begin{document}

\title{Coulomb energy of uniformly-charged spheroidal shell systems}
\author{Vikram Jadhao}
\email{vjadhao1@jhu.edu}
\altaffiliation[Present address: ]{Department of Physics and Astronomy, Johns Hopkins University, Baltimore, MD, 21218}
\affiliation{Department of Materials Science and Engineering, Northwestern University, Evanston, Illinois 60208, USA}
\author{Zhenwei Yao}
\affiliation{Department of Materials Science and Engineering, Northwestern University, Evanston, Illinois 60208, USA}
\author{Creighton K. Thomas}
\affiliation{Department of Materials Science and Engineering, Northwestern University, Evanston, Illinois 60208, USA}
\author{Monica Olvera de la Cruz}
\email{m-olvera@northwestern.edu}
\affiliation{Department of Materials Science and Engineering, Northwestern University, Evanston, Illinois 60208, USA}
\affiliation{Department of Chemistry, Northwestern University, Evanston, Illinois 60208, USA}
\affiliation{Department of Chemical and Biological Engineering, Northwestern University, Evanston, Illinois 60208, USA}
\affiliation{Department of Physics and Astronomy, Northwestern University, Evanston, Illinois 60208, USA}

\begin{abstract} 
We provide exact expressions for the electrostatic energy of uniformly-charged prolate and oblate spheroidal shells. We find that uniformly-charged prolate spheroids of eccentricity greater than 0.9 have lower Coulomb energy than a sphere of the same area. For the volume-constrained case, we find that a sphere has the highest Coulomb energy among all spheroidal shells. Further, we derive the change in the Coulomb energy of a uniformly-charged shell due to small, area-conserving perturbations on the spherical shape. Our perturbation calculations show that buckling-type deformations on a sphere can lower the Coulomb energy.
Finally, we consider the possibility of counterion condensation on the spheroidal shell surface. We employ a Manning-Oosawa two-state model approximation to evaluate the renormalized charge and analyze the behavior of the equilibrium free energy as a function of the shell's aspect ratio for both area-constrained and volume-constrained cases. Counterion condensation is seen to favor the formation of spheroidal structures over a sphere of equal area for high values of shell volume fractions.
\end{abstract}

\maketitle

\section{Introduction}
Shapes of physical systems as diverse as blood cell membranes, colloidal
particles, nanowires, and galaxies are often considered as spheroidal with
varying degrees of eccentricity. Many charged structures such as
colloids or emulsions are usually modeled as spheroidal shells 
with a uniform surface charge density. 
To our best knowledge, the expression for the Coulomb energy of a 
uniformly-charged spheroidal shell is not available in the literature. 
In this article, we provide the needed result. Further, we derive a general expression for the
change in the Coulomb energy of a uniformly-charged shell due to 
small, area-conserving perturbations on the spherical shape. Using the result, 
we explore the existence of deformations that can lower the
electrostatic energy relative to the unperturbed charged sphere.

We note that a closed-form expression for the electrostatic potential energy of a solid
homogeneously-charged spheroid can been obtained \cite{landau.fieldtheory}.
Further, the electrostatic energy of a conducting spheroidal shell as a
function of the aspect ratio is available elsewhere \cite{landau.electrodynamics}. 
Calculations of the electrostatic potential for a system of point charges inside dielectric
spheroids \cite{shaozhong.prolate,shaozhong.oblate} have been performed as well. 
On the other hand, it is useful to note the work in the context of solving Poisson-Boltzmann equation in spheroidal geometry \cite{hsu,hsu.linearpb}.
In this paper, we provide a comprehensive study of the homogeneously-charged spheroidal shell system which 
has been missing in the literature. 

Primary motivations behind our calculations stem from the 
study carried out in Ref.~\onlinecite{jto}, where equilibrium shapes of charged, soft shells constrained to maintain a fixed volume were analyzed using molecular dynamics simulations. Some of the results derived in the present article were employed to verify the oblate-shaped shell structures found in Ref.~\onlinecite{jto} and calculate the effects of ion condensation on the equilibrium shape of these structures. The supporting information associated with Ref.~\onlinecite{jto} also contained a brief derivation of the electrostatic energy of a uniformly-charged oblate spheroidal shell. In this paper, we derive the general expression for the Coulomb energy of prolate spheroidal shells. For the sake of completeness we also include the derivation of the Coulomb energy of oblate shells showing details that were omitted in Ref.~\onlinecite{jto}. We analyze the variation of the Coulomb energy of spheroidal shells, subject to the constraints of fixed area or volume, as the aspect ratio of the shell is 
changed. We also examine the effects of ion condensation, computed via a Manning-Oosawa two-state model analysis \cite{manning,oosawa}, on the variation of the equilibrium free energy of the shell-counterion system. Finally, we note that it is straightforward to augment the energy expressions obtained here to reveal the gravitational potential energy of a uniformly-dense spheroidal surface which is often used as a model to study galaxies \cite{galactic.dynamics}.

The key findings of this paper are: 
i) A homogeneous prolate (cigar-shaped)
spheroidal shell with eccentricity greater than $\sim$ 0.9 has a lower
electrostatic energy than a spherical shell of the same area. The lowest-energy
shape of the shell, constrained to maintain its area, is a very long and thin prolate spheroid whose energy
approaches zero as its major-axis length is stretched to infinity. 
ii) For shells that are constrained to maintain their volume, the spherical shape has
the maximum Coulomb energy. An infinitely long and thin wire-like shape and a
thin, flat disc of infinite area, are the degenerate lowest-energy shapes with
vanishing energy. 
iii) Perturbation calculations show that, for the case of
fixed area constraint, the Coulomb energy of a uniformly-charged sphere can be lowered by a buckling-type deformation. 
iv) Results from the two-state model approximation of the shell-counterion system show that counterion condensation favors the formation of spheroidal structures over a sphere of equal area for high values of shell volume fractions.

The paper is organized as follows. In Sec.~\ref{sec:energy.spheroidal}, we provide the expression for the Coulomb energy of a uniformly-charged spheroidal shell, discuss the important limiting cases, and specialize the expression for the case of constant area and constant volume constraints. Sec.~\ref{sec:generic.deformation} shows the comparison between the energy of a sphere and nearly-spherical structures formed by a small, generic perturbation around the spherical shape. In Sec.~\ref{sec:charge.renormalization}, we discuss the effects of charge renormalization on the energies obtained for the spheroidal shell system and Sec.~\ref{sec:conclusion} is the conclusion. Appendices A and B present the derivation of the Coulomb energy of uniformly-charged prolate and oblate spheroidal shells respectively, and in Appendix C we derive the electrostatic energy of a uniformly-charged circular disc.

\section{Electrostatic energy of uniformly-charged spheroidal shells}\label{sec:energy.spheroidal}
\begin{figure}
\centerline{\includegraphics[scale=0.2]{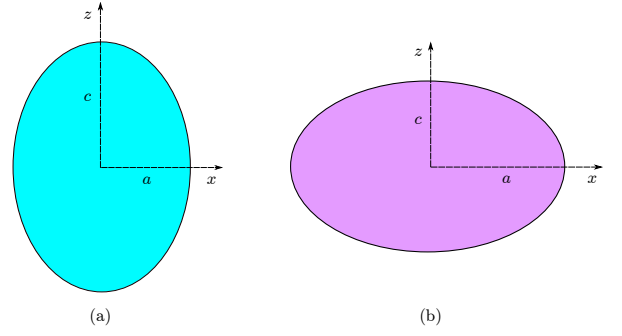}}
\caption
{
Cross-section of a spheroid normal to the $y$-axis with $c$ and $a$ as the semi-axis lengths and $\lambda = c/a$ being the spheroid's aspect ratio. (a) Prolate spheroid with $\lambda > 1$. (b) Oblate spheroid with $\lambda < 1$.
} 
\label{spheroid}
\end{figure}

Consider a spheroidal shell with charge $Q$ distributed uniformly over its surface such that the charge density is given by $\sigma = Q / A$, where $A$ is the area of the spheroid. A spheroid is an ellipsoid two of whose semi-principal axes are equal. Assuming that the equal lengths correspond to the dimensions along the $x$ and $y$ axes, such that the cross section normal to the $z$-axis is a circle, we can describe the spheroidal shell via the equation: 
\beq
\frac{r^2}{a^2} + \frac{z^2}{c^2} = 1,
\eeq
where $a$ and $c$ are the semi-principal axes, and $r=\sqrt{x^2+y^2}$ is the distance between the point on the surface of the spheroid and the $z$-axis. A prolate spheroid is a spheroid where $c > a$, whereas an oblate spheroid corresponds to the $c < a$ condition (see Fig.~\ref{spheroid}). Clearly when $a = c$, the spheroid reduces to a sphere. 

It is convenient to characterize the spheroid by defining the aspect ratio $\lambda$ defined as 
\beq\label{eq:aspect.ratio}
\lambda = \frac{c}{a}.
\eeq
Values of $\lambda < 1$ correspond to oblate spheroid whereas a prolate spheroid is associated with $\lambda > 1$. $\lambda \to 0$ corresponds to a circular disc, $\lambda = 1$ is a sphere, and $\lambda \to \infty$ limit produces an infinitely long and thin rod-like spheroid. It is also useful to introduce the eccentricity $e_{\textrm{p}}$ of a prolate spheroid defined as:
\beq
e_{\textrm{p}} = \sqrt{1 - \frac{a^2}{c^2}}.
\eeq
Similarly, we have the eccentricity $e_{\textrm{o}}$ for an oblate spheroid:
\beq
e_{\textrm{o}} = \sqrt{1 - \frac{c^2}{a^2}}. 
\eeq
Note that either eccentricities lie between 0 and 1. When $e_{\textrm{o}}, e_{\textrm{p}}\to0$, the spheroid reduces to a sphere. The limit $e_{\textrm{o}} \to 1$ leads to a circular disc and $e_{\textrm{p}} \to 1$ corresponds to a very long and thin rod like shape. We will invoke these limits at several places in what follows.

In Appendix \ref{sec:prolate.energy}, we derive the electrostatic energy of a uniformly-charged prolate spheroidal shell. Our calculations, which employ the standard method of separation of variables \cite{jackson,lebedev}, lead to the following result:
\beq
\begin{split}
U(e_{\textrm{p}},c,\sigma) &= 4\pi^2\sigma^2 c^3 \frac{1-e_{\textrm{p}}^2}{e_{\textrm{p}}} \, \times \\
&\sum_{n\in \textrm{even}} \frac{2n+1}{2} \,  P_{n}(1/e_{\textrm{p}}) 
Q_n(1/e_{\textrm{p}}) \left(H_{n}(e_{\textrm{p}})\right)^2,
\end{split}
\eeq
where $n$ is an even integer, $P_n$ and $Q_n$ are Legendre functions of the first and second kind respectively, and $H_n(e_{\textrm{p}})$ is the integral 
\beq
H_{n}(e_{\textrm{p}}) = \int_{0}^{\pi} \sqrt{1 - e_{\textrm{p}}^{2} \textrm{cos}^{2} v} \,P_n(\textrm{cos}v) \textrm{sin}v \, dv.
\eeq
It is useful to express the result in terms of the total charge $Q$ rather than $\sigma$. Using the fact that the area of a prolate spheroid is 
\beq\label{eq:p.area}
\mathscr{A}_{\textrm{p}}(e_{\textrm{p}},c) = 2\pi c^2\sqrt{1-e_{\textrm{p}}^2}\,T(e_{\textrm{p}}),
\eeq
where
\beq
T(e_{\textrm{p}}) = \sqrt{1-e_{\textrm{p}}^2} + \frac{\textrm{sin}^{-1}(e_{\textrm{p}})}{e_{\textrm{p}}},
\eeq
we obtain the electrostatic energy of a homogeneously-charged prolate spheroidal shell to be:
\beq\label{eq:p.U}
U_{\textrm{p}} = \frac{Q^2}{2c\,e_{\textrm{p}}T(e_{\textrm{p}})^2}
\sum_{n\in \textrm{even}} (2n+1)\, P_{n}\left(\frac{1}{e_{\textrm{p}}}\right)  Q_n\left(\frac{1}{e_{\textrm{p}}}\right) H_{n}(e_{\textrm{p}})^2
\eeq
where $U_{\textrm{p}}\equiv U_{\textrm{p}}(e_{\textrm{p}},c,Q)$.
Eq.~\eqref{eq:p.U} provides the first key result of this paper. We point out that the above expression for the Coulomb energy is obtained assuming that the prolate spheroidal shell is in vacuum. If the medium surrounding the shell is polarizable and uniform, the above result for the energy must be scaled down by the dielectric constant of the medium.

Appendix \ref{sec:oblate.energy} provides the derivation for the Coulomb energy of the oblate spheroidal shell. We note that a brief account of this derivation appears in the supplementary information of Ref.~\onlinecite{jto}. We find the electrostatic energy to be:
\beq
\begin{split}
&U(e_{\textrm{o}},a,\sigma) = \frac{4\pi^2\sigma^2 a^3i}{e_{\textrm{o}}} \, \times \\
&\sum_{n\in \textrm{even}} \frac{2n+1}{2} \,  P_{n}\left(i\, \frac{\sqrt{1-e_{\textrm{o}}^2}}{e_{\textrm{o}}}\right) 
Q_n\left(i\,\frac{\sqrt{1-e_{\textrm{o}}^2}}{e_{\textrm{o}}}\right) I_{n}(e_{\textrm{o}})^2,
\end{split}
\eeq
where $n$ is an even integer and $I_n(e_{\textrm{o}})$ is the integral 
\beq
I_{n}(e_{\textrm{o}}) = \int_{0}^{\pi} \sqrt{1 - e_{\textrm{o}}^{2} \textrm{sin}^{2} v} \,P_n(\textrm{cos}v) \textrm{sin}v \, dv.
\eeq
Once again, we express below the result in terms of the total charge $Q$. Using the fact that the area of an oblate spheroid is 
\beq\label{eq:o.area}
\mathscr{A}_{\textrm{o}}(e_{\textrm{o}},a) = 2\pi a^2 S(e_{\textrm{o}}),
\eeq
where
\beq
S(e_{\textrm{o}}) = 1 + \left(\frac{1}{e_{\textrm{o}}} - e_{\textrm{o}}\right) \textrm{tanh}^{-1} e_{\textrm{o}},
\eeq
we obtain the electrostatic energy of the uniformly-charged oblate spheroidal shell to be:
\beq\label{eq:o.U}
\begin{split}
&U_{\textrm{o}} = \frac{Q^2i}{2ae_{\textrm{o}}S(e_{\textrm{o}})^2} \times \\
&\sum_{n\in \textrm{even}} (2n+1) \, 
\, P_{n}\left(i \frac{\sqrt{1-e_{\textrm{o}}^2}}{e_{\textrm{o}}}\right)  Q_n\left(i \frac{\sqrt{1-e_{\textrm{o}}^2}}{e_{\textrm{o}}}\right) I_{n}(e_{\textrm{o}})^2,
\end{split}
\eeq
where $U_{\textrm{o}}\equiv U_{\textrm{o}}(e_{\textrm{o}},a,Q)$. Eq.~\eqref{eq:o.U} provides the second key result of this paper.
In several physical situations, geometric constraints such as the constraint of fixed area or volume are naturally present and it is of interest to find the shell shape that minimizes the Coulomb energy when only area-preserving or volume-conserving deformations are allowed. 
Using the above expressions for the electrostatic energy, we analyze the Coulomb energy of a shell that is subjected to these constraints and present the results in sections \ref{sec:aconstraint} and \ref{sec:vconstraint}. Before that we take a quick look at the limiting cases of the energy expressions found in Eqs.~\eqref{eq:p.U} and \eqref{eq:o.U}.

\subsection{Limiting cases}
We recall that both the prolate and oblate eccentricities lie between 0 and 1. First we let $e_{\textrm{p}},e_{\textrm{o}}$ approach zero, which corresponds to a spherical shell, and find
\beq\label{eq:sphereenergy}
U_{\textrm{p}}(e_{\textrm{p}}\to0,a,Q) = U_{\textrm{o}}(e_{\textrm{o}}\to0,a,Q) = \frac{Q^2}{2a}.
\eeq
We recover the well-known result for the energy of a uniformly-charged spherical shell. 

Taking the limit $e_{\textrm{o}}\to1$ of the oblate energy expression in Eq.~\eqref{eq:o.U} gives
\beq\label{eq:eto1}
U_{\textrm{o}}(e_{\textrm{o}}\to1,a,Q;n\le6) = 0.84872\frac{Q^2}{a},
\eeq
where, because of the rapid convergence of the sum in Eq.~\eqref{eq:o.U}, we have included terms up to $n=6$ in obtaining the above result. The limit $e_{\textrm{o}}\to1$ corresponds to the shape of a circular disc. Unlike the spherical case, an exact expression for the energy of a uniformly-charged disc to our best knowledge has not been reported in the literature. It is possible to arrive at this energy starting from the electrostatic potential on the surface of the disc derived in Ref.~\cite{ciftja}. We show the derivation of the energy in Appendix \ref{app:disc}. The result is 
\beq\label{eq:discenergy}
U_{\textrm{disc}} = \frac{8}{3\pi}\frac{Q^2}{a},
\eeq
where $a$ is the radius of the disc and $Q = \pi a^2 \sigma$ is the total charge.
We can now compare the disc energy obtained in Eq.~\eqref{eq:eto1} with the above exact result and find the deviation to be $\sim 0.01\%$.
Clearly, the two energies are in very good agreement. 

The prolate energy expression, Eq.~\eqref{eq:p.U}, in the limit of prolate eccentricity approaching unity gives $U_{\textrm{p}}(e_{\textrm{p}}\to1,a,Q) = \infty$, that is, the energy diverges. In this limit, the prolate spheroid is transformed into a thin, long rod-like shape where the width of the rod is shrunk at the same time as the length of the rod is stretched out. The divergence of the energy arises because, as the width narrows, the distance between the charges on the surface shrinks faster in comparison with the charges growing apart due to the extension in the length. As we will find out in the next section, when we impose constraints of constant shell area or volume, the prolate energy no longer diverges when the aforesaid limit is taken. 

\subsection{Charged spheroidal shells of equal area}\label{sec:aconstraint}
We begin with the results for the application of area constraint. We will assume the reference shell shape in the analysis to be the sphere. 
The area of a prolate spheroid is given by Eq.~\eqref{eq:p.area} which we rewrite below:
\beq\label{eq:p.area2}
\mathscr{A}_{\textrm{p}}(e_{\textrm{p}},c) = 2\pi c^2\sqrt{1-e_{\textrm{p}}^2}
\left(\sqrt{1-e_{\textrm{p}}^2} + \frac{\textrm{sin}^{-1}(e_{\textrm{p}})}{e_{\textrm{p}}} \right).
\eeq
This equation suggests that if $\mathscr{A}$ is fixed, $c$ and $e_{\textrm{p}}$ are coupled. Assuming that the area is constrained to that of a sphere of radius $R$, $c$ and $e_{\textrm{p}}$ become related via the equation:
\beq\label{eq:cA.ep}
c(e_{\textrm{p}},R) = R\sqrt{\frac{2}{\sqrt{1-e_{\textrm{p}}^2}\left(\sqrt{1-e_{\textrm{p}}^2} + \frac{\textrm{sin}^{-1}(e_{\textrm{p}})}{e_{\textrm{p}}}\right)}}.
\eeq
Eliminating $c$ from Eq.~\eqref{eq:p.U} using the above equation, we arrive at the expression for the electrostatic energy of a uniformly-charged prolate spheroidal shell constrained to a fixed area of $4\pi R^2$ as a function of the eccentricity $e_{\textrm{p}}$:
\begin{eqnarray}\label{eq:p.UA}
U_{\textrm{p},A}(e_{\textrm{p}},R,Q) &=& \frac{Q^2 \sqrt{
1-e_{\textrm{p}}^2 + \sqrt{\frac{1}{e_{\textrm{p}}^2}-1}\,\textrm{sin}^{-1}(e_{\textrm{p}})}}{2R\sqrt{2}e_{\textrm{p}}T(e_{\textrm{p}})^2}\times\\
&&\!\!\!\!\!\!\!\!
\sum_{n\in\textrm{even}} (2n+1)\, P_{n}\left(1/e_{\textrm{p}}\right)  Q_n\left(1/e_{\textrm{p}}\right) H_{n}(e_{\textrm{p}})^2.\nonumber
\end{eqnarray} 

We proceed similarly with the case of an oblate spheroidal shell. The area of an oblate spheroid is given by Eq.~\eqref{eq:o.area} which we rewrite below:
\beq
\mathscr{A}_{\textrm{o}}(e_{\textrm{o}},a) = 2\pi a^2 
\left(1 + \left(\frac{1}{e_{\textrm{o}}} - e_{\textrm{o}}\right) \textrm{tanh}^{-1} e_{\textrm{o}} \right),
\eeq
where $\textrm{tanh}^{-1}$ denotes the inverse hyperbolic tangent function. We note that in the limit $e_{\textrm{o}}\to 1$, the oblate shell reduces to a structure resembling a circular disc having two faces with total area $2\pi a^2$.
If $\mathscr{A}_{\textrm{o}}$ is fixed, $e_{\textrm{o}}$ and $a$ are coupled and recalling that the area is constrained to the value $4\pi R^2$, we obtain the relation:
\beq\label{eq:aA.eo}
a(e_{\textrm{o}},R) = R\sqrt{\frac{2}{1 + \left(\frac{1}{e_{\textrm{o}}} - e_{\textrm{o}}\right) \textrm{tanh}^{-1} e_{\textrm{o}}}}.
\eeq
Eliminating $a$ from Eq.~\eqref{eq:o.U} using the above equation, we arrive at the expression for the electrostatic energy of a uniformly-charged oblate spheroidal shell subject to the constraint of fixed area:
\bea\label{eq:o.UA}
&&U_{\textrm{o},A}(e_{\textrm{o}},R,Q) = \frac{Q^2i\sqrt{1 + \left(\frac{1}{e_{\textrm{o}}} - e_{\textrm{o}}\right) \textrm{tanh}^{-1} e_{\textrm{o}}}}{2R\sqrt{2} e_{\textrm{o}}S(e_{\textrm{o}})^2} \times \\
&&\sum_{n\in\textrm{even}} (2n+1) \, 
\, P_{n}\left(i \frac{\sqrt{1-e_{\textrm{o}}^2}}{e_{\textrm{o}}}\right)  Q_n\left(i \frac{\sqrt{1-e_{\textrm{o}}^2}}{e_{\textrm{o}}}\right) I_{n}(e_{\textrm{o}})^2.\nonumber
\eea

It is useful to express the energies in Eqs.~\eqref{eq:p.UA} and \eqref{eq:o.UA} as a function of the aspect ratio $\lambda$ defined in Eq.~\eqref{eq:aspect.ratio}. Noting that $e_{\textrm{p}} = \sqrt{\lambda^2 - 1}/\lambda$ and $e_{\textrm{o}} = \sqrt{1-\lambda^2}$, we arrive at the result: 
\begin{equation}\label{eq:po.UAlambda}
U_{A}(\lambda) = \left\{ 
\begin{array}{ll}
U_{\textrm{o},A}(\sqrt{1 - \lambda^2},R,Q) & 0 < \lambda < 1 \\
U_{\textrm{p},A}(\sqrt{\lambda^2 - 1}/\lambda,R,Q) & \lambda \ge 1,
\end{array} \right.
\end{equation}
where we have suppressed the dependence of $U_A$ on other variables for brevity. The values of $\lambda \ge 1$ correspond to prolate spheroids and $0 < \lambda < 1$ region corresponds to oblate spheroids. Equation \eqref{eq:po.UAlambda} provides the Coulomb energy of uniformly-charged spheroidal shells, all having the same area, for values of the aspect ratio $\lambda$ ranging from 0 to $\infty$. 

\begin{figure}
\centerline{\includegraphics[scale=0.7]{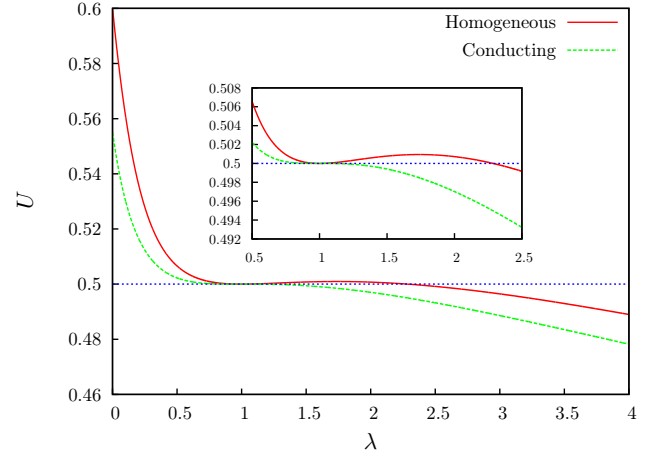}}
\caption
{
Coulomb energy of charged spheroidal shells of equal area as a function of their aspect ratio $\lambda$. The region $0 \le \lambda < 1$ corresponds to oblate shells and values of $\lambda > 1$ represent prolate shells. Solid red line corresponds to the homogeneously-charged spheroidal shell and dashed green line is the result for the conducting spheroidal shell. The dotted blue line is energy of a spherical shell (reference) which is 0.5 for either cases (homogeneous or conducting). The above results are for $Q = 1$, $R = 1$. See text for the meaning of symbols.
} 
\label{afix}
\end{figure}
We now analyze the variation of $U_A(\lambda)$ as $\lambda$ is changed. We set $Q=R=1$ for simplicity. We first confirm that $U_A(\lambda \to 1) = 0.5$ as should be the case for a uniformly-charged spherical shell ($\lambda = 1 \Rightarrow a = c$) for the aforesaid parameters. Further, it is easy to check that $U_A(\lambda \to 0) \sim 0.6$ which is equivalent to the Coulomb energy of an infinitely thin, uniformly-charged circular disc of unit radius. Taking the opposite limit, we find $U_A(\lambda \to \infty) = 0$, which suggests that the Coulomb energy of a very thin and long wire-like shape (of finite area) vanishes. We graph the function $U_A(\lambda)$ as a function of $\lambda$ in Fig.~\ref{afix}. We choose the number of terms appearing in the series expansion in the energy expressions derived in Eqs.~\eqref{eq:p.UA} and \eqref{eq:o.UA} to be $n = 6$ as the series converges rapidly. As is evident from Fig.~\ref{afix}, we observe that the sphere shape is a local minimum. However, as the aspect ratio 
is increased to values beyond $\sim 2.286$, which corresponds to a prolate eccentricity of $\sim 0.9$, the Coulomb energy is lowered below that of the sphere. The energy continues to decrease as $\lambda$ increases further and we find that the shape that corresponds to the lowest Coulomb energy is the very thin and long prolate spheroidal shell (of area $4\pi$), the minimum energy being 0. Further, the energy of the oblate spheroidal shell increases upon increasing its eccentricity (or lowering the aspect ratio $\lambda$), with the thin circular disc corresponding to the shape of maximum energy.

It is instructive to compare the results for the case of uniformly-charged spheroidal shells with that of conducting spheroidal shells. Exact expressions for the latter have been obtained elsewhere \cite{landau.electrodynamics}. Using these exact results, the particular expression for the case where the area is held fixed can be easily derived and we summarize the final results below:
\begin{equation}\label{eq:po.CUAlambda}
\mathscr{U}_{A}(\lambda) = \left\{ 
\begin{array}{ll}
\frac{Q^2}{2a\left(\sqrt{1-\lambda^2},R\right)}\frac{1}{\sqrt{1-\lambda^2}}\textrm{tan}^{-1}\frac{\sqrt{1-\lambda^2}}{\lambda}  \quad 0 < \lambda < 1 \\
\frac{Q^2}{2c\left(\sqrt{1-\lambda^2}/\lambda,R\right)}\frac{\lambda}{\sqrt{\lambda^2-1}}\textrm{tanh}^{-1}(\frac{\sqrt{\lambda^2-1}}{\lambda}) \quad \lambda \ge 1.
\end{array} \right.
\end{equation}
In the above equation, the functions $a$ and $c$ are given by Eqs.~\eqref{eq:aA.eo} and \eqref{eq:cA.ep} respectively. Similar to the above analysis for homogeneously charged shells, we can evaluate the variation of $\mathscr{U}_{A}$ as a function of $\lambda$ and we obtain the green dashed line in Fig.~\ref{afix}. We find that this line is always below the red solid line. This implies that allowing the surface charges to move freely, as is the case with the conducting shell, lowers the Coulomb energy. In addition, it is important to note that all the prolate conducting shells have a lower energy than a spherical conducting shell. In sharp contrast, for uniformly-charged shells, the spherical shape is a clear local minimum as evidenced by the red solid curve in the inset of Fig.~\ref{afix}.

\subsection{Charged spheroidal shells of equal volume}\label{sec:vconstraint}
We now analyze the Coulomb energy of spheroidal shells that are subjected to the volume constraint, that is all the shells have the same volume. This analysis is very similar to the one presented in the last subsection and so we will keep the following discussion brief.
The volume of a prolate spheroid is 
\beq\label{eq:vol.prolate}
\Omega_{\textrm{p}}(e_{\textrm{p}},c) = \frac{4}{3}\pi c^3 \left(1 - e_{\textrm{p}}^2\right).
\eeq
This equation suggests that if $\Omega_{\textrm{p}}$ is fixed to $(4/3)\pi R^3$, then $c$ and $e_{\textrm{p}}$ are related via the equation:
\beq\label{eq:cV.ep}
c(e_{\textrm{p}},R) = \frac{R}{\left(1-e_{\textrm{p}}^2\right)^{1/3}}.
\eeq
Eliminating $c$ from Eq.~\eqref{eq:p.U} using the above equation, we arrive at the expression for the electrostatic energy of a uniformly-charged prolate spheroidal shell constrained to a fixed volume of $(4/3)\pi R^3$ as a function of the eccentricity $e_{\textrm{p}}$:
\begin{eqnarray}\label{eq:p.UV}
U_{\textrm{p},V}(e_{\textrm{p}},R,Q) &=& \frac{\left(1-e_{\textrm{p}}^2\right)^{1/3} Q^2}{2R\,e_{\textrm{p}}T(e_{\textrm{p}})^2}\times\\
&&\!\!\!\!\!\!\!\!
\sum_{n\in\textrm{even}} (2n+1)\, P_{n}\left(1/e_{\textrm{p}}\right)  Q_n\left(1/e_{\textrm{p}}\right) H_{n}(e_{\textrm{p}})^2\nonumber.
\end{eqnarray} 

We proceed similarly with the case of an oblate spheroidal shell. The volume of an oblate spheroid is given by 
\beq\label{eq:vol.oblate}
\Omega_{\textrm{o}}(e_{\textrm{o}},a) = \frac{4}{3}\pi a^3 \sqrt{1 - e_{\textrm{o}}^2}.
\eeq
If $\Omega_{\textrm{o}}$ is fixed, $e_{\textrm{o}}$ and $a$ are coupled and recalling that the volume is constrained to the value $(4/3)\pi R^3$, we obtain the relation:
\beq\label{eq:aV.eo}
a(e_{\textrm{o}},R) = \frac{R}{\left(1-e_{\textrm{o}}^2\right)^{1/6}}.
\eeq
Eliminating $a$ from Eq.~\eqref{eq:o.U} using the above equation, we arrive at the expression for the electrostatic energy of a uniformly-charged oblate spheroidal shell subject to the constraint of fixed volume:
\beq\label{eq:o.UV}
\begin{split}
&U_{\textrm{o},V}(e_{\textrm{o}},R,Q) = \frac{Q^2i(1-e_{\textrm{o}}^2)^{1/6}}{2Re_{\textrm{o}}S(e_{\textrm{o}})^2} \times \\
&\sum_{n\in\textrm{even}} (2n+1) \, 
\, P_{n}\left(i \frac{\sqrt{1-e_{\textrm{o}}^2}}{e_{\textrm{o}}}\right)  Q_n\left(i \frac{\sqrt{1-e_{\textrm{o}}^2}}{e_{\textrm{o}}}\right) I_{n}(e_{\textrm{o}})^2.
\end{split}
\eeq

\begin{figure}
\centerline{\includegraphics[scale=0.7]{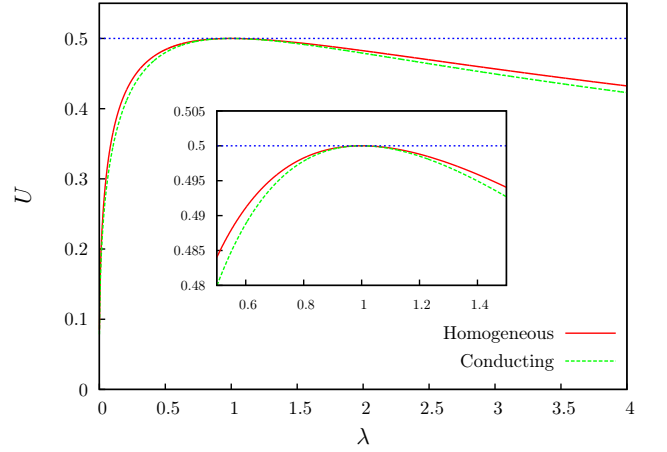}}
\caption
{
Coulomb energy of charged spheroidal shells of equal volume as a function of their aspect ratio $\lambda$. The region $0 \le \lambda < 1$ corresponds to oblate shells and values of $\lambda > 1$ represent prolate shells. Solid red line corresponds to the homogeneously-charged spheroidal shell and dashed green line is the result for the conducting spheroidal shell. The dotted blue line is energy of a spherical shell (reference) which is 0.5 for either cases (homogeneous or conducting). The above result is for $Q = 1$, $R = 1$. See text for the meaning of symbols.
} 
\label{vfix}
\end{figure}

As in the case of the area constraint, it is useful to express the energies in Eqs.~\eqref{eq:p.UV} and \eqref{eq:o.UV} as a function of the aspect ratio $\lambda$. Carrying out the transformation from eccentricities $e_{\textrm{p}}, e_{\textrm{o}}$ to $\lambda$, we arrive at the result: 
\begin{equation}\label{eq:po.UVlambda}
U_{V}(\lambda) = \left\{ 
\begin{array}{ll}
U_{\textrm{o},V}(\sqrt{1 - \lambda^2},R,Q) & 0 < \lambda < 1 \\
U_{\textrm{p},V}(\sqrt{\lambda^2 - 1}/\lambda,R,Q) & \lambda \ge 1,
\end{array} \right.
\end{equation}
where we have suppressed the dependence of $U_V$ on other variables for brevity. Equation \eqref{eq:po.UVlambda} provides the Coulomb energy of uniformly-charged spheroidal shells, all having the same volume, for values of $\lambda$ ranging from 0 to $\infty$. 

In Fig.~\ref{vfix}, we plot the change in $U_V$ as $\lambda$ is varied for the parameters $Q = 1$ and $R = 1$. We find that $U_V(\lambda \to 1) = 0.5$ as expected for the homogeneously-charged sphere. We notice that the sphere represents the shape of maximum Coulomb energy which is in sharp contrast with the above analyzed case of the area-constrained charged shells. We observe in Fig.~\ref{vfix} that every other energy value is degenerate, in the sense that there are two \emph{distinct} shapes that correspond to the same energy, one of the shapes being an oblate and the other a prolate. We find that the minimum energy for the volume-constrained charged spheroidal shell is 0. The two shapes that correspond to this value are a thin charged disc (of infinite area), and a long and thin prolate spheroidal shape (of infinite area). Again, this is in sharp contrast with the area-constrained charged shell where the shape of minimum energy is unique and is a prolate as pointed out in Sec.~\ref{sec:aconstraint}.

We compare the results for the uniformly-charged shells with conducting spheroidal shells. The energy expression for the volume-constrained conducting shell problem is the same as Eq.~\eqref{eq:po.CUAlambda} except that the functions $a$ and $c$ are now determined by Eqs.~\eqref{eq:aV.eo} and \eqref{eq:cV.ep} respectively. Similar to Fig.~\ref{afix}, we find the Coulomb energy of a conducting shell is always lower than that of the homogeneously charged shell. Again, we attribute this to the fact that for a conducting shell, the charges will move until the shell surface becomes an equipotential, and this movement always lowers the Coulomb energy. Finally, we note that in Fig.~\ref{vfix}, the energy profiles for the two different scenarios (homogeneous and conducting) are similar in shape which is in contrast with Fig.~\ref{afix} for the area-constrained case. 

\section{Energy analysis of perturbed spheres}\label{sec:generic.deformation}
Results of the preceding sections indicate that a uniformly-charged sphere is locally stable to perturbations towards a prolate or oblate spheroid if the deformations preserve the surface area. The stability of the charged sphere for arbitrary perturbations in shape that preserve the area and maintain the uniformity of the surface charge distribution is not addressed yet. In this section, we explore this problem by deriving the expression for the energy variation of the shell due to small generic perturbations on the spherical shape.

Consider a sphere of radius $R_0$ that is represented by 
\beq
\vec{R}_0(\theta,\phi)=(R_0 \sin\theta \cos\phi, R_0 \sin\theta \sin\phi,
R_0\cos\theta),
\eeq 
where $\theta \in [0, \pi]$ and $\phi \in [0,2\pi)$.
A generic small perturbation $h(\theta,\phi)$ on the sphere produces a shape that can be represented by 
\begin{eqnarray}\label{R_ansatz}
\vec{R}(\theta,\phi)=\vec{R}_0(\theta,\phi) +
h(\theta,\phi)\hat{N}(\theta,\phi), \label{R_pert}
\end{eqnarray} 
where $\hat{N}(\theta,\phi)=R_0^{-1}\vec{R}_0(\theta,\phi)$ is the unit normal vector on the sphere.
Note that $\vec{R}(\theta,\phi)=\vec{R}_0(\theta,\phi)(1+h(\theta,\phi)/R_{0})$ and we consider perturbations to be small when they satisfy the condition $|h|/R_0 \ll 1$.
The metric tensor of the undeformed sphere is
\beq
g_{ij} = \left( \begin{array}{cc}
R_0^2 & 0 \\
0 & R_0^2\sin^2\theta
\end{array} \right),
\eeq
with its determinant $g_0=(g_0)_{11} (g_0)_{22} = R_0^4\sin^2\theta$.
The determinant of the metric tensor of the deformed shape defined in
Eq.(\ref{R_pert}) is~\cite{zhong1989bending}:
\be\label{g}
g=g_0+\delta g,
\ee 
where 
\be\label{dg} 
\delta g=g_0
\left[4hH+(g_0)^{ij}h_i h_j +h^2 (4H^2+2K)\right]+ {\cal O}(h^3),
\ee 
with 
$H=1/R_0$ and $K=(1/R_0)^2$.

We assume that the charges on the surface interact via a pair potential $V(r)$
which only depends on the distance $r$ between them. We consider a homogeneous
surface and allow only those deformations that keep the area unchanged. Thus,
the charge density is the same even in the deformed shape. The charge elements
on the unperturbed shape are $dq_0=\sigma_0 dA_0 = \sigma_0\sqrt{g_0}d^2\vec{x}$ 
where $\sigma_0 = Q/A_0$ is the uniform charge density. 
The charge elements on the perturbed shape are 
$dq=\sigma_0 dA = \sigma_0\sqrt{g}d^2\vec{x}$. 
Note that $d^2\vec{x}$ is a shorthand for $d\theta d\phi$. The change in the interaction energy brought about by the deformation of the charged sphere is 
\begin{equation}\label{H}
\begin{split}
&H[h]=\frac{\sigma_0^2}{2}\int \sqrt{g(\vec{x})}
d^2\vec{x} \sqrt{g(\vec{x}')} d^2\vec{x}'
\,V\left(\left|\vec{R}(\vec{x}')-\vec{R}(\vec{x})\right| \right) \\
&-\frac{\sigma_0^2}{2}\int \sqrt{g_0(\vec{x})} d^2\vec{x} \sqrt{g_0(\vec{x}')} d^2\vec{x}' 
\,V\left(\left|\vec{R}_0(\vec{x}')-\vec{R}_0(\vec{x})\right| \right),
\end{split}
\end{equation}
where the integration is over $x_1=\theta,\ x_2=\phi,\ x'_1=\theta',\ x'_2=\phi'$. 
We note that if $H < 0$, the energy of the perturbed
sphere is lower than that of the spherical system. The
distance between two arbitrary points on the deformed sphere is 
\bea
r&=&\left|\vec{R}(\vec{x}')-\vec{R}(\vec{x})\right| \nonumber \\
&=&\left|\left(\vec{R}_0(\vec{x}')-\vec{R}_0(\vec{x}) \right) +\left( h(\vec{x}')\hat{N}(\vec{x}')
-h(\vec{x})\hat{N}(\vec{x})\right)\right| \nonumber \\ 
&=& \left|\vec{r}_0 +\delta\vec{r}\right| = r_0 +\Delta r,\label{r} 
\eea where
$\vec{r}_0=\vec{R}_0(\vec{x}')-\vec{R}_0(\vec{x})$ and 
$\delta \vec{r} = h(\vec{x}')\hat{N}(\vec{x}') - h(\vec{x})\hat{N}(\vec{x})$. 
The $r$ in Eq.~\eqref{r} can be expanded in terms of $\delta\vec{r}$ according to 
\be \left| \vec{f}+\delta \vec{f} \,\right| =
f+\Delta f , \ee where $f=|\vec{f}\,|$ and 
\be\label{Deltaf} \Delta
f=\frac{\vec{f}\cdot \delta\vec{f}}{f} +\frac{1}{2}\frac{\delta \vec{f}\cdot \delta \vec{f}}{f}
-\frac{1}{2}\frac{(\vec{f}\cdot\delta \vec{f})^2}{f^3}+{\cal O}(\delta f^3).
\ee 
Therefore, 
\bea\label{Deltar2} 
\Delta r&=&\frac{1}{2}\frac{r_0}{R_0}\left( h(\vec{x})+h(\vec{x}')
\right) + \frac{1}{2r_0}\left(h(\vec{x})-h(\vec{x}')\right)^2\\
&+&\frac{1}{2}\frac{r_0}{R_0^2}h(\vec{x})h(\vec{x}') 
- \frac{r_0}{8R_0^2}\left(h(\vec{x})+h(\vec{x}')\right)^2+{\cal O}(h^3).\nonumber 
\eea

We expand $V(r)$ and $\sqrt{g(\vec{x})g(\vec{x}')}$ in terms of $h$ up to the quadratic order: 
\bea\label{Vr}  
V(r) &=& V(r_0+\Delta r)\\
&=&V(r_0)+\frac{d V(r_0)}{dr_0}
\Delta r + \frac{1}{2} \frac{d^2 V(r_0)}{dr_0^2} (\Delta r)^2 +{\cal O}(\Delta
r^3), \nonumber 
\eea 
\bea\label{gg}
&&\sqrt{g(\vec{x})}\sqrt{g(\vec{x}')}=\sqrt{g_0 g_0'}\times \\ &&\left[ 1+\frac{1}{2}\left(
\frac{\delta g}{g_0}+ \frac{\delta g'}{g'_0}+\frac{\delta g \delta g'}{g_0
g'_0} \right) -\frac{1}{8}\left( \frac{\delta g}{g_0}+ \frac{\delta
g'}{g'_0}\right)^2 \right]+{\cal O}(h^3),\nonumber 
\eea 
where the prime on functions $g_0$ and $\delta g$ denotes that the variable of the function is $\vec{x}'$,
{\cal{e.g.}, $g'_0$ is short for $g_0(\vec{x}')$. Note
that $\delta g$ contains terms linear in $h$.

We formally write $\sqrt{g(\vec{x})}\sqrt{g(\vec{x}')}=\sqrt{g_0 g'_0} \left(1+\alpha(h)+\beta(h^2)
\right)$ and $V(r)=V(r_0)+A(h)+B(h^2)$, where $\alpha(h)$ and $A(h)$ are terms
linear in $h$ and $\beta(h^2)$ and $B(h^2)$ represent terms quadratic in $h$.
Eq.~\eqref{H} can therefore be written as 
\bea\label{H2}
&&H=\frac{\sigma_0^2}{2}\int d^2\vec{x} d^2\vec{x}' \sqrt{g_0g'_0} \left[
A(h)+\alpha(h)V(r_0)\right]\nonumber\\ &&+\frac{\sigma_0^2}{2}\int d^2\vec{x} d^2\vec{x}'
\sqrt{g_0g'_0}\left[ B(h^2)+ \alpha(h)A(h) +\beta(h^2) V(r_0)\right] \nonumber\\
&&+ {\cal O}(h^3).  \eea 
Using Eqs.~\eqref{dg} and \eqref{Deltar2}, we obtain
$A(h)=\frac{r_0}{2R_0}\frac{dV(r_0)}{dr_0}(h+h')$ and
$\alpha(h)=\frac{2}{R_0}(h+h')$.  Up to the linear term in $h$, Eq.~\eqref{H2}
becomes 
\bea\label{H1} 
H = \frac{\sigma_0^2R_0^3}{2} \int d\theta d\phi &&d\theta'
d\phi'\sin\theta \sin\theta' \left[2V(r_0)+\frac{r_0}{2}\frac{dV(r_0)}{dr_0}
\right] \nonumber\\
&&\times\left(h(\theta,\phi)+h(\theta',\phi') \right) + {\cal O}(h^2).  
\eea 
Note: 
\bea r_0&&(\theta,\phi, \theta',\phi') = \\
&&R_0\sqrt{2\left[1-\cos\theta\cos\theta'-\cos(\phi-\phi')\sin\theta\sin\theta'\right]}.\nonumber
\eea

The existence of terms linear in $h$ in Eq.~\eqref{H1} suggests that the spherical shape is not necessarily an energy
extreme, opening the possibility of shapes with lower energy than that of the
sphere. For $V(r_0)= 1/ r_0^{\alpha}$, the term in the square bracket becomes
\beq\label{vdv} 2V(r_0)+\frac{1}{2}r_0
\frac{dV(r_0)}{dr_0}=\frac{1}{r_0^{\alpha}}\left(2-\frac{\alpha}{2}\right).  
\eeq 
We find that for $\alpha = 1$, which represents the Coulomb potential, this term is
non-zero. This indicates that the spherical system may be unstable to the long-range
Coulomb interaction. We note that for $\alpha = 4$, the above term vanishes implying
there are no linear terms in $h$ for this potential. Furthermore, the sign of the
energy changes as the value of $\alpha$ goes beyond 4, implying the distinct
stability behaviors of the spherical shape for long- and short-range potentials.

We now investigate if there exists a perturbation that can lower the Coulomb
energy of the uniformly-charged sphere, in other words, a deformation for which $H<0$.
We consider a generic form for the perturbation $h$ represented by
$h(\theta,\phi)= \sum_{l,m} a_{lm}Y_{lm}(\theta,\phi)$, where $Y_{lm}$ are the spherical harmonic functions with $l=0,1,2,\ldots$ and $m=-l,-l+1,\ldots,l$. Here $a_{lm}$ are the unknown expansion coefficients or modes. As $h$ is taken to be a small perturbation, we require $|a_{lm}|\ll R_{0}$. We take $m=0$, thus examining axisymmetric deformations. 
The constraint of fixed area leads to the relation between the coefficients
$a_{l0}$. The variation of area \cite{zhong1989bending} is 
\beq\label{dA} 
\delta A=2\sqrt{4\pi} a_{00}R_0 + \sum_{lm} |a_{lm}|^2\left(1+\frac{1}{2}l(l+1)
\right).  \eeq Setting $\delta A=0$ leads to 
\beq \label{eq:al0.constraint}
2\sqrt{4\pi} a_{00} R_0 =-\sum_{lm} |a_{lm}|^2 \left(1+\frac{1}{2}l(l+1)\right).  
\eeq 
We find that $a_{00}$ is always negative, implying a uniform shrinking of the shell to
preserve the area. When only uniform shrinking (or expansion) is allowed, $a_{l0}=0$ for $l>0$. For this case, from Eq.~\eqref{eq:al0.constraint}, we obtain the relation: $2\sqrt{4\pi} a_{00} R_0 + a_{00}^2 = 0$. This equation has two solutions, $a_{00} = -4\sqrt{\pi} R_0$, which is unphysical as $|a_{00}| > R_0$, and $a_{00}=0$. The latter solution indicates no deformations on the sphere.

Now we consider perturbations characterized by two modes $a_{00}$ and
$a_{10}$: 
\beq\label{eq:h2modes}
h(\theta, \phi) = \frac{a_{00}}{\sqrt{4 \pi}} + a_{10}
\sqrt{\frac{3}{4\pi}} \cos\theta,
\eeq
where $\theta \in [0,\pi]$. 
We note that the modes $a_{00}$ and $a_{10}$ are coupled due to the constraint of fixed area, Eq.~\eqref{eq:al0.constraint}, leading to the relation: 
\beq\label{a01}
\frac{1}{2}a_{00}^2 + 2\sqrt{\pi}R_0a_{00} + a_{10}^2 = 0. 
\eeq
Recall that $a_{00} \le 0$ and the above equation implies $a_{10}$ can assume positive or negative values.
Physically, the perturbation $h$ of Eq.~\eqref{eq:h2modes} corresponds to a buckling of the spherical shape.
If we identify $\theta=0$ as the north pole 
and consequently $\theta=\pi$ as the south pole, we find that the north pole is buckled inward for $a_{10}<0$ and the south pole is buckled inwards for $a_{10}>0$.

\begin{figure}[h] 
\centerline{\includegraphics[scale=0.7]{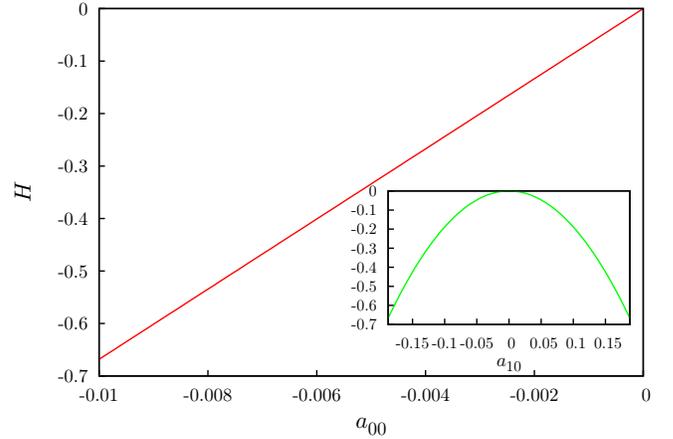}}
\caption{
The total Coulomb energy change $H$ vs the coefficient $a_{00}$.
The inset shows the variation of $H$ with $a_{10}$. The shape of the
perturbed sphere is characterized by the two modes $a_{00}$ and $a_{10}$,
which are coupled due to the constraint of the fixed area.
} 
\label{H_a00}
\end{figure}

The change in the total Coulomb energy $H$ of the system (up to linear terms in $h$) for this perturbation can be calculated from Eq.~\eqref{H1} and we find:
\beq\label{eq:Hc1}
H =  \frac{3\sigma_0^2R_0^3}{2\sqrt{4\pi}} \int d\theta d\phi d\theta'd\phi'
\sin\theta \sin\theta'\frac{a_{00} + a_{10}\sqrt{3}\cos\theta}{r_0}.
\eeq
We note that the shape of the perturbed sphere is independent 
of the sign of the coefficient $a_{10}$; to reverse the sign of $a_{10}$ 
is to rotate the shape by $\pi$. This suggests that terms linear in $a_{10}$ should be absent in $H$.
By making use of the parity of the integrand in the four quadrants: $\{\theta\in [0,\pi/2],\theta'\in [0,\pi/2]\}$, $\{\theta\in [0,\pi/2],\theta'\in [\pi/2,\pi]\}$, $\{\theta\in [\pi/2,\pi],\theta'\in [0,\pi/2]\}$, and $\{\theta\in [\pi/2,\pi],\theta'\in [\pi/2,\pi]\}$, we find that the integral involving the $a_{10}\cos\theta$ term in Eq.~\eqref{eq:Hc1} vanishes. Noting that the remaining integral (involving the $a_{00}$ term) in Eq.~\eqref{eq:Hc1} can be read as the total Coulomb energy of a uniformly-charged spherical shell, we arrive at the following analytical result for $H$:
\beq\label{eq:Hc2}
H = 12\pi^{3/2}\sigma_0^2R_0^2a_{00}.
\eeq
Because $a_{00}\le 0$, we see from Eq.~\eqref{eq:Hc2} that $H\le 0$. In other words, the perturbation in Eq.~\eqref{eq:h2modes} lowers the Coulomb energy of the original unperturbed spherical shape. In Fig.~\ref{H_a00}, we plot $H$ vs  $a_{00}$ for $\sigma_0 = 1$ and $R_0=1$. Recalling that $a_{00}$ characterizes the amount of buckling in the deformed shape, we find that the system energy keeps decreasing with the rise in the buckling of the shape. Noting that $a_{00}$ and $a_{10}$ are coupled via Eq.~\eqref{eq:al0.constraint}, we obtain $H$ as a function of $a_{10}$ using Eq.~\eqref{eq:Hc2}:
\beq
H = 12\pi^{3/2}\sigma_0^2R_0^2\left(\sqrt{4\pi R_0^2-2a_{10}^2}-\sqrt{4\pi R_0^2}\right).
\eeq
As expected, $H$ is found to be independent of the sign of $a_{10}$. In the inset of Fig.~\ref{H_a00}, we show the variation of $H$ with $a_{10}$. The above calculations demonstrate that a uniformly-charged sphere is electrostatically unstable to a buckling-type deformation in the constraint of fixed area.

Alongside the computation of $H$, which quantifies the change in the total Coulomb energy, it is instructive to examine how the Coulomb energy changes locally at specific points on the shell surface as a result of the deformation proposed in Eq.~\eqref{eq:h2modes}. 
We define the local Coulomb energy at a point as the interaction energy of the charge element at that point with all the other charges on the surface. For the uniformly-charged sphere (undeformed state), the local Coulomb energy is the same at all points on the surface. 

We discuss, without any loss of generality, the case of $a_{10}<0$ for which the north pole ($\theta=0$) is buckled inwards. For simplicity, we perform calculations for two points on the surface: the north and the south poles. For the perturbation $h$ given by Eq.~\eqref{eq:h2modes}, we find the change in the local Coulomb energy (relative to the original spherical conformation) of a charge element located at the north pole to be:
\beq 
dH = \frac{1}{2}\sigma_0^2 dA'  \int
R_0 \sin\theta d\theta d\phi \frac{3}{2r_0} \left(h\left(\theta,\phi\right) +
h\left(0,0\right)\right), \label{dH} 
\eeq 
where $dA'=\sqrt{g_{0}(x_{N})}d^{2}x_{N}$ is the area element associated with the charge in the undeformed (spherical) conformation and $r_0 = 2R_0\sin(\theta/2)$. We are primarily interested in the sign of $dH$; if this interaction energy is negative (positive), that
would imply the local Coulomb energy for the deformed shape is lowered (raised) relative to the original sphere. We define 
\beq\label{HN} 
H_N = \sigma_0^2 R_0\int d\theta d\phi \sin\theta
\frac{3}{2r_0} (h\left(\theta,\phi\right) + h\left(0,0\right)) 
\eeq 
and note that $H_N$ and $dH$ have the same sign and differ by a constant prefactor.
Upon substituting $h$ from Eq.~\eqref{eq:h2modes} in Eq.~\eqref{HN}, the latter becomes 
\beq \label{H_N}
H_N = 6\sqrt{\pi}\sigma_0^2
\left(a_{00} + \frac{2}{\sqrt{3}}a_{10}\right).
\eeq
We follow a similar procedure to obtain the change in the local Coulomb energy density $H_S$ at the south pole. We find that $H_S$ is given by Eq.~\eqref{H_N} with $a_{10}$ replaced by $-a_{10}$. Finally, noting that the two modes $a_{00}$ and $a_{10}$ are coupled via Eq.~\eqref{eq:al0.constraint}, we express $H_N$ and $H_S$ as functions of $a_{00}$:
\beq
H_{N,S} = 6\sqrt{\pi}\sigma_0^2
\left(a_{00} \mp \sqrt{\frac{2}{3}\left(-4\sqrt{\pi}a_{00}R_0^2- a_{00}^2\right)} \right),
\eeq
where the $-(+)$ sign corresponds to the north (south) pole. We plot $H_N$ and $H_S$ vs  $a_{00}$ in Fig.~\ref{E_northpole} for $\sigma_0=1, R_0=1$. We find that as the magnitude $a_{00}$ of the deformation (buckling) is increased, the local Coulomb energy is lowered at the north pole ($H_N \le 0$). On the other hand, at the south pole, the local Coulomb energy is higher relative to its value in the undeformed spherical case ($H_S \ge 0$).
\begin{figure}[h] 
\centerline{\includegraphics[scale=0.7]{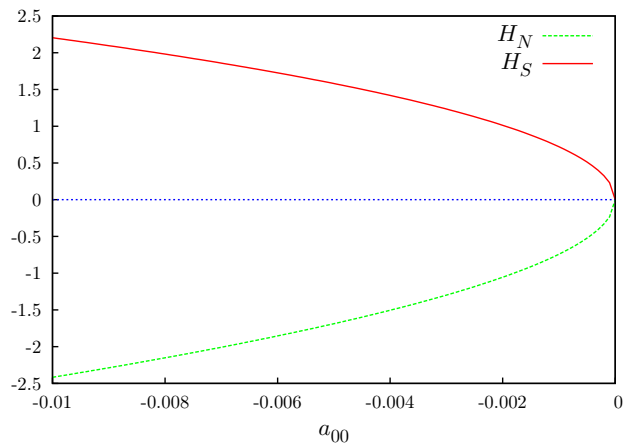}}
\caption{$H_N$ (dashed green) and $H_S$ (solid red) vs $a_{00}$ for the perturbation $h$ characterized by two modes $a_{00}$ and $a_{10}$. The above plot is for the case when $a_{10} < 0$, which corresponds to the inward buckling of the north pole. See text for the meaning of the symbols.}
\label{E_northpole} 
\end{figure}

\section{Charge renormalization in spheroidal shells}\label{sec:charge.renormalization}
In previous sections, we analyzed the Coulomb energy of uniformly-charged spheroidal and sphere-like shells and determined the conformations that correspond to the lowest Coulomb energy under a given geometric constraint. The shell was considered to be an isolated system in vacuum with charges embedded on its surface. In realistic settings, however, we expect the shells to be in an environment where the surrounding medium (solvent) contains counterions that neutralize the shell charge, rendering the overall system electroneutral. In this light, results obtained thus far are reliable in the event that the associated counterions remain in the bulk, far from the shell surface. In situations where a significant fraction of the total number of counterions condense on the shell, the free energy of the shell-counterion system, determines the equilibrium shell conformations. We note that at infinite dilution, in the spherical case, the entropy is expected to dominate the shell-counterion Coulomb attraction, leading 
to no counterion condensation on the shell surface. However, at finite shell concentrations (volume fractions), condensation is expected to occur, even in salt-free settings \cite{alexander,borukhov,monica}. The condensation of the counterions can be viewed as renormalizing the (bare) charge on the shell \cite{alexander}. This renormalized charge and consequently the behavior of the equilibrium free energy of the shell-counterion system can be obtained in a qualitative way by using the Manning-Oosawa two-state model \cite{oosawa,manning,borukhov,gillespie}. In this section, we use this two-state model approximation and compute the renormalized charge on uniformly-charged oblate and prolate spheroidal shells at finite shell concentrations in salt-free settings and find the variation of the equilibrium free energy of the system as a function of the shell aspect ratio.

We investigate the effects of counterion condensation on spheroidal shells of equal area and spheroidal shells of equal volume. For either case, we take the area (volume) to be constrained to that of a sphere of radius $R$. We consider a Wigner-Seitz (WS) cell of volume $V_{\textrm{WS}}$ containing a single shell of volume $\Omega$, with $Q$ charge on its surface, placed at the center. We work with finite values of the shell concentration $\eta = \Omega/V_{\textrm{WS}}$. The cell also contains $N$ counterions, each of charge $Q/N$ making the overall shell-counterion system electroneutral. The counterions are separated into two distinct groups: free ions and condensed ions. The condensed counterions are restricted to have translational motion in a thin layer of volume $V_c = A(\lambda,R) b$ surrounding the shell, where $A(\lambda,R)$ is the area of the shell and $b$ is the thickness of the layer. Note that as we restrict our analysis to shells constrained to a fixed area or volume, the area $A$ (and the volume $\Omega$) of the shell can be considered as a function of the aspect ratio $\lambda$ and $R$. Free ions occupy the available space in the WS cell which in the dilute limit can be approximated to be the volume of the cell. We choose experimentally relevant parameters: total charge $Q=600$ electron units (which amounts to $\sim$100 mV of surface potential), $N=1000$ counterions, and $R=10$ nm. Calculations are performed for room temperature $T=300$ K and we take water as the dielectric medium surrounding the shell.   

Let $\alpha$ be the fraction of counterions that condense. Clearly, $(1-\alpha)N$ ions remain free in the bulk. Further, the condensed ions neutralize the surface charge on the shell reducing the net charge to 
$(1-\alpha)Q$. We approximate the WS cell to be spherical with volume $V_{\textrm{WS}} = (4/3)\pi R_{\textrm{WS}}^3$, with $R_{\textrm{WS}}$ being the radius of the cell that gets determined by the shell volume fraction (concentration) $\eta$. We vary the shell concentration $\eta$ from $10^{-12}$ to $10^{-4}$. We write the free energy (in units of $k_BT$) associated with the shell as:
\begin{equation}\label{eq:F}
\begin{split}
F(\alpha,\lambda) &= (1-\alpha)^2 U(\lambda) + 
\alpha N \, \textrm{ln} \left(\frac{\alpha N \Lambda^3}{A(\lambda,R)b}\right) - \alpha N \\ 
&+ (1-\alpha)N \, \textrm{ln} \left( \frac{(1-\alpha)N\Lambda^3}{V_{\textrm{WS}}} \right)
- (1-\alpha)N,
\end{split}
\end{equation}
where $\Lambda$ is the thermal de Broglie wavelength. The first term is the electrostatic potential energy of the shell with renormalized charge $(1-\alpha)Q$. The function $U$ (shown below) represents the bare (unrenormalized) Coulomb energy and is determined based on particular geometric constraint employed. 
The second and third terms stem from the entropic contribution of the $\alpha N$ condensed ions, and the last two terms correspond to the entropy of $(1-\alpha) N$ free counterions. Note that within this model, the entropy of both free and condensed ions is assumed to be that of an ideal gas. Also, just like $U$, the form of the area $A$ of the shell and its volume $\Omega$ depends on the constraint applied as shown below.

For the case of shells subject to the equal area constraint, the function $U(\lambda)$ reads
\begin{equation}\label{eq:po.UAlambda2}
U(\lambda) = U_{A}(\lambda) = \left\{ 
\begin{array}{ll}
l_{\textrm{B}}U_{\textrm{o},A}(\sqrt{1 - \lambda^2},R,Q) & 0 < \lambda < 1 \\
l_{\textrm{B}}U_{\textrm{p},A}(\sqrt{\lambda^2 - 1}/\lambda,R,Q) & \lambda \ge 1,
\end{array} \right.
\end{equation}
where $U_{\textrm{o},A}$ and $U_{\textrm{p},A}$ are available from Eq.~\eqref{eq:po.UAlambda}, which provides the expression for the Coulomb energy of uniformly-charged spheroidal shells of equal area, and the above result is expressed in units of $k_BT$ by introducing the Bjerrum length $l_{\textrm{B}}$. 
Recall that the values of $\lambda \ge 1$ correspond to prolate spheroids and $0 < \lambda < 1$ region corresponds to oblate spheroids. 
Owing to the constraint, the area of the shell is simply $A(\lambda,R) = 4\pi R^2$. The shell volume $\Omega$ follows from Eqs.~\eqref{eq:vol.prolate}, \eqref{eq:cA.ep}, \eqref{eq:vol.oblate}, and \eqref{eq:aA.eo}:
\begin{equation}\label{eq:shell.volume}
\Omega(\lambda, R) = \left\{ 
\begin{array}{ll}
\Omega_{\textrm{o}}\left(\sqrt{1 - \lambda^2},a\left(\sqrt{1 - \lambda^2},R\right)\right) & \!\!\!\!\! 0 < \lambda < 1 \\
\Omega_{\textrm{p}}\left(\sqrt{\lambda^2 - 1}/\lambda,c\left(\sqrt{\lambda^2 - 1}/\lambda,R\right)\right)&  \lambda \ge 1.
\end{array} \right.
\end{equation}

For the volume-constrained problem, $U(\lambda)$ follows from  Eq.~\eqref{eq:po.UVlambda}: 
\begin{equation}\label{eq:po.UVlambda2}
U(\lambda) = U_{V}(\lambda) = \left\{ 
\begin{array}{ll}
l_{\textrm{B}}U_{\textrm{o},V}(\sqrt{1 - \lambda^2},R,Q) & 0 < \lambda < 1 \\
l_{\textrm{B}}U_{\textrm{p},V}(\sqrt{\lambda^2 - 1}/\lambda,R,Q) & \lambda \ge 1.
\end{array} \right.
\end{equation}
Following Eqs.~\eqref{eq:p.area}, \eqref{eq:o.area}, \eqref{eq:cV.ep}, and \eqref{eq:aV.eo}, the area function $A$ for this case becomes
\begin{equation}\label{eq:shell.area}
A(\lambda, R) = \left\{ 
\begin{array}{ll}
\mathscr{A}_{\textrm{o}}\left(\sqrt{1 - \lambda^2},a\left(\sqrt{1 - \lambda^2},R\right)\right) & \!\!\!\!\! 0 < \lambda < 1 \\
\mathscr{A}_{\textrm{p}}\left(\sqrt{\lambda^2 - 1}/\lambda,c\left(\sqrt{\lambda^2 - 1}/\lambda,R\right)\right) & \lambda \ge 1.
\end{array} \right.
\end{equation}
Finally, owing to the constraint, the shell volume is simply $\Omega = 4\pi R^3/3$.

We approximate the thickness $b$ of the condensed layer by the Gouy-Chapman (GC) length $b = 1/(2\pi l_{\textrm{B}}\sigma)$, where $\sigma$ is the unrenormalized charge density on the shell surface \cite{gillespie}. 
Higher charge density or longer Bjerrum length leads to a stronger shell-counterion attraction implying a thin  condensed layer; this is indeed reflected when $b$ is chosen as the layer thickness as seen from the above expression. We also note that the GC length is a length scale associated with the planar interface and hence our analysis is limited to the regime where $b$ is shorter than the characteristic lengths associated with the shell. We have carried out the following analysis by choosing the Bjerrum length $l_{\textrm{B}}$ as the thickness of the condensed-layer and we find no changes in the conclusions reached below.

The free energy $F$ in Eq.~\eqref{eq:F} can be considered as a function of $\lambda$ and $\alpha$. For a given $\lambda$ (shape), we minimize the free energy with respect to $\alpha$ to find the fraction of counterions that condense on the shell. We obtain the extremum condition:
\begin{equation}\label{eq:alpha}
-\xi (1 - \alpha) + N \,\textrm{ln} \left( \frac{\alpha}{1-\alpha} \frac{1}{\eta} \frac{\Omega(\lambda,R)}{A(\lambda,R)b}
\right) = 0,
\end{equation}
where $\xi = 2 U(\lambda)$ measures the strength of the Coulomb interactions and $\eta$ is the volume fraction of the shells given by
\beq
\eta = \frac{\Omega}{V_{\textrm{WS}}}.
\eeq
For a given $\lambda$ and $\eta$, we solve Eq.~\eqref{eq:alpha} using \emph{Mathematica} and obtain $\alpha$ as a function of $\lambda$. We carry out the study for a wide range of shell volume fractions ranging from $10^{-12}$ to $10^{-4}$. Using the value of $\alpha$, the renormalized electrostatic energy $\overline{U}$ of the shell at equilibrium is known from Eq.~\eqref{eq:po.UAlambda2} or Eq.~\eqref{eq:po.UVlambda2} (depending on the constrained problem under study), by replacing $Q$ with $(1-\alpha)Q$: 
\beq\label{eq:Uateqm}
\overline{U}(\lambda,Q) = U(\lambda,(1-\alpha)Q).
\eeq
Employing the above result and the equilibrium value of the condensate fraction $\alpha$ (obtained as the solution of Eq.~\eqref{eq:alpha}), it is easy to show that the  difference in the equilibrium free energies of a spheroidal shell and a spherical shell, $dF$, defined as 
\beq
dF(\lambda) = F(\lambda) - F(\lambda\to 1)
\eeq
is given by:
\beq\label{eq:dFateqm}
dF = \frac{1+\alpha}{1-\alpha} \overline{U} - \frac{1+\alpha_{\textrm{s}}}{1-\alpha_{\textrm{s}}} \overline{U}_{\textrm{s}} + N\textrm{ln}\frac{1-\alpha}{1-\alpha_{\textrm{s}}}.
\eeq
In Eq.~\eqref{eq:dFateqm}, $\alpha_{\textrm{s}}$ and $\overline{U}_{\textrm{s}}$ denote, respectively, the values of the condensate fraction and the renormalized Coulomb energy for a spherical shell. 

\begin{figure} \centerline{\includegraphics[scale=0.7]{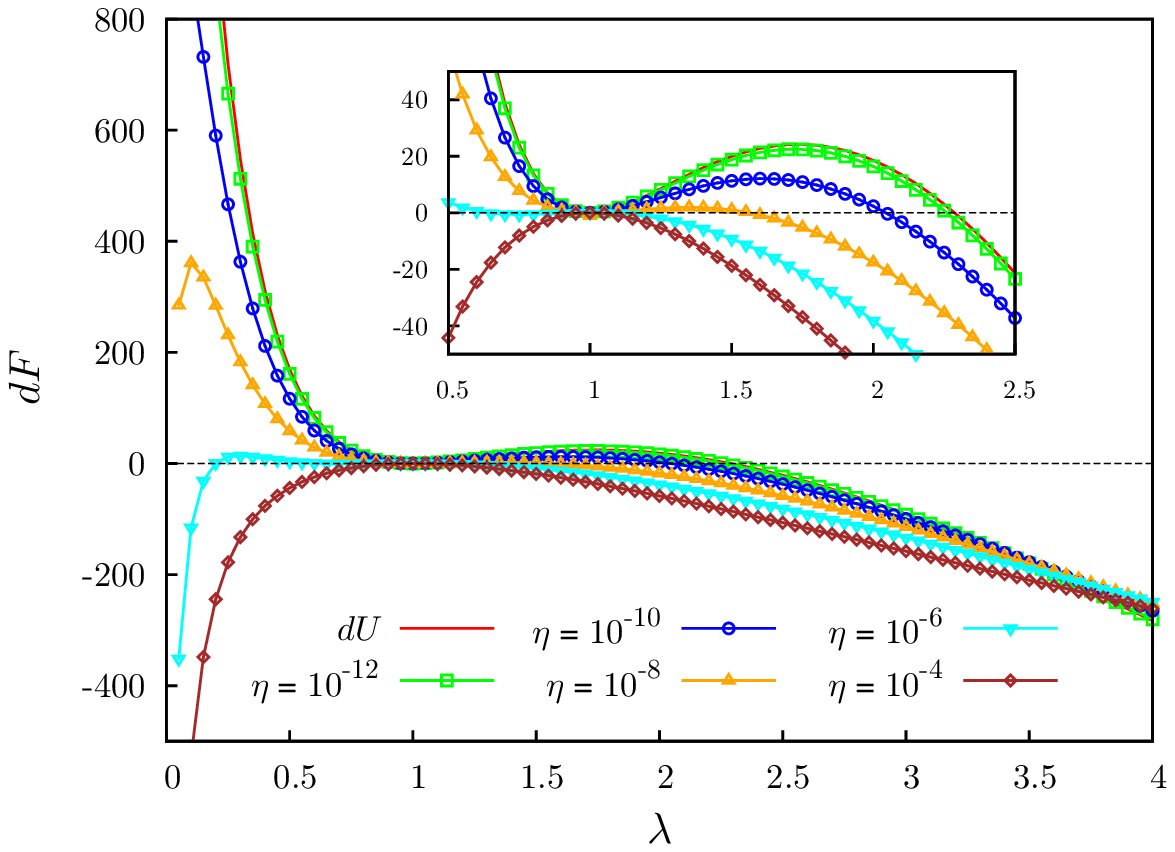}}
\caption
{
Equilibrium free energy difference, $dF$, between a homogeneously-charged spheroidal shell and a sphere of identical parameters ($z = 0.6$, $N=1000$, $R = 10$ nm) as a function of the aspect ratio $\lambda$ for the area-constrained system. The effects of charge renormalization due to counterion condensation are included via a two-state model analysis calculation. The red curve is the case for the counterion-free system and acts as a reference line. All other curves take into account ion condensation on the shell surface and correspond to different values of the shell volume fraction $\eta$. Results are shown for $\eta=10^{-12}$ (green squares), $10^{-10}$ (blue circles), $10^{-8}$ (orange triangles), $10^{-6}$ (cyan inverted triangles), and $10^{-4}$ (brown diamonds). We find that the spherical conformation, which is a local minimum for the isolated spheroidal system (red line), becomes a free-energy maximum at $\eta = 10^{-4}$. 
} 
\label{afix.dF}
\end{figure}

\begin{figure} \centerline{\includegraphics[scale=0.7]{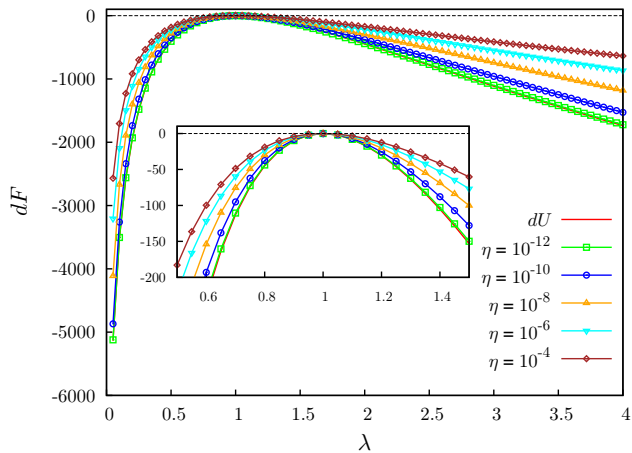}}
\caption
{
Equilibrium free energy difference, $dF$, between a homogeneously-charged spheroidal shell and a sphere of identical parameters ($z = 0.6$, $N=1000$, $R = 10$ nm) as a function of the aspect ratio $\lambda$ for the volume-constrained system. The effects of charge renormalization due to counterion condensation are included via a two-state model analysis calculation. The red curve is the case for the counterion-free system and acts as a reference line. All other curves take into account ion condensation on the shell surface and correspond to different values of the shell volume fraction $\eta$. Results are shown for $\eta=10^{-12}$ (green squares), $10^{-10}$ (blue circles), $10^{-8}$ (orange triangles), $10^{-6}$ (cyan inverted triangles), and $10^{-4}$ (brown diamonds). We find that in the event of ion condensation, for all values of $\eta$, the spherical conformation continues to have the highest free energy among all spheroidal shapes. 
} 
\label{vfix.dF}
\end{figure}

We now analyze the variation of $dF$ as $\lambda$ is changed. We consider $\lambda$ values from 0 to 4 like in the study of the Coulomb energy of isolated, homogeneously-charged spheroidal shells recorded in Figs.~\ref{afix} and \ref{vfix}. We begin with the case of fixed shell area. In Fig.~\ref{afix.dF}, we plot $dF$, computed from Eq.~\eqref{eq:dFateqm}, as a function of $\lambda$ for various values of $\eta$.  The red solid line is the Coulomb energy $dU$ of isolated spheroidal shells of equal area measured relative to the Coulomb energy of the spherical shell with identical parameters. This line acts as a reference curve to all other lines which are the result of taking ion condensation into consideration. Recall that for the counterion-free case, the spherical shape is a local energy minimum as evidenced by the $dU$ plot. For low $\eta$ $(= 10^{-12})$, which corresponds to a very dilute system, we find that the $dF$ curve (green squares) lies in the vicinity of the no-condensation result ($dU$). However, as 
$\eta$ rises, we observe significant deviations from the unrenormalized energy curve for both oblate ($\lambda < 1$) and prolate $\lambda > 1$ regions. 

We find that for all shapes ($\lambda$), as the volume fraction $\eta$ is increased, $\alpha$ increases, that is more counterions condense on the shell surface. A major consequence of the enhanced charge renormalization is the reduction of the positive free energy difference between the spheroidal shell and the sphere (see Fig.~\ref{afix.dF} inset). Further, for high $\eta$ values ($\eta = 10^{-6}$, $10^{-4}$), we find that all prolate shells have less free energy as compared to the sphere which is in stark contrast from the no-condensation result (red line). For the same $\eta$ values, we also find that oblate shells with small aspect ratios have lower free energy than a spherical shell ($dF < 0$). For the volume fraction of $\eta = 10^{-4}$ (brown diamonds), the spherical shell has the maximum equilibrium free energy among all shapes. Thus, according to the above analysis based on the two-state model, for the area-constrained shell system, counterion condensation has a profound effect in modifying the 
energy landscape associated with the isolated shell, favoring the formation of spheroidal structures over the spherically-shaped ones as the shell volume fraction is increased. 

Fig.~\ref{vfix.dF} shows the variation of $dF$ with $\lambda$ for different $\eta$ values in the case of fixed-volume constraint. Once again, the solid red line corresponds to the counterion-free system and is the Coulomb energy of the uniformly-charged spheroidal shell measured relative to the electrostatic energy of the sphere. As Fig.~\ref{vfix} shows, the spherical shape is the conformation of maximum energy for this particular constraint. This conclusion remains unchanged when we include the effects of charge renormalization via the two-state model analysis as seen from Fig.~\ref{vfix.dF}. For a very dilute system, $\eta = 10^{-12}$ (green squares), we find that the $dF$ curve lies in the vicinity of the $dU$ line. Increasing $\eta$ leads to a rise in $\alpha$ which is seen to weaken the (negative) difference between the equilibrium free energies of spheroidal and spherical shells (see Fig.~\ref{vfix.dF} inset). Thus, judging by the variation of $dF$ determined by the two-state model analysis, we find that 
oblate and prolate shaped structures continue to be energetically favored over the spherical conformation in the event of counterion condensation. Further, the monotonic trend of free energy decrease with increasing the eccentricity of the spheroidal shell observed for the isolated shell system is seen to persist in the wake of counterion condensation as well. 

We note that as the counterions are mobile, under certain conditions, the charged shell is better approximated as a conducting surface as opposed to a homogeneously-charged one. In that event, the above analysis can be carried out using the expressions of electrostatic energy for conducting spheroidal shells provided in Eq.~\eqref{eq:po.CUAlambda}. 
We observe that for all values of $\eta$, the conducting spheroidal shell has a lower free energy upon deformation in comparison with the homogeneously-charged shell (see Ref.~\onlinecite{jto} for details of this calculation for oblate spheroidal shells under the constraint of constant volume).
Further, we find that the main conclusions regarding the effects of counterion condensation reached above for either constraints remain unchanged when we repeat the two-state model analysis assuming that the shell is an equipotential surface.  

Finally, we note that the study of condensation effects based on the above two-state model employs a number of approximations. For example, the free energy associated with this model, Eq.~\eqref{eq:F}, does not take into account the shell-counterion and counterion-counterion Coulomb interactions explicitly. Another approximation is to put the shell in a counterion-only, salt-free environment. Also, the distribution of counterions around the spheroidal shell is considered to be isotropic which is clearly a simplification for shapes that deviate significantly from the spherical conformation. In this light, we view the above results as qualitative. Quantitative results that address many of the aforesaid simplifications can be obtained by employing approaches based on the solution of the Poisson-Boltzmann equation for spheroidal geometry \cite{alvarez}. 
 
\section{Conclusion}\label{sec:conclusion}
We report the exact expression for the electrostatic energy of a uniformly-charged spheroidal shell. We  analyze the variation in the electrostatic energy as the aspect ratio of the shell is changed from $0$ to $\infty$ for the area-constrained and volume-constrained cases. The prolate spheroidal shell with its major-axis length stretched to infinity is found to have the lowest Coulomb energy among spheroidal shells of equal area. Further, we reveal the non-monotonous variation in the Coulomb energy when a spherical shell is elongated to a prolate spheroid keeping the shell area fixed. For spheroidal shells that have the same volume, a sphere has the highest Coulomb energy. In addition, our perturbation calculations show that there exist area-conserving buckling-type deformations on the sphere that can lower the total Coulomb energy. For the spheroidal shell system, we use a two-state model of free and condensed ions to evaluate the renormalization of the shell charge due to counterion condensation. We find 
that ion condensation 
has a significant effect in modifying the free energy landscape with spheroidal structures being favored over a sphere of the same area as the shell volume fraction is increased. These results add to the theoretical foundation required to understand the control of spheroidal shapes in materials using electrostatics in combination with other forces such as those arising due to the elastic nature of the material \cite{jto}.

\section{Acknowledgments}
V. J. thanks Francisco J. Solis for insightful discussions. We gratefully acknowledge the financial support from the Office of the Director of Defense Research and Engineering (DDR\&E) and the Air Force Office of Scientific Research (AFOSR) under Award No. FA9550-10-1-0167 and the Office of Basic Energy Sciences within Department of Energy Grant DE-FG02-08ER46539. 

\appendix
\section{Coulomb energy of uniformly-charged prolate spheroidal shells}\label{sec:prolate.energy}
In this section we derive the expression shown in Eq.~\eqref{eq:p.U} for the Coulomb energy of a homogeneously-charged prolate spheroidal shell. We consider prolate spheroidal coordinates $u, v, \phi$, which are related to the Cartesian coordinates $x,y,z$ by
\bea
x &=& ce_{\textrm{p}}\, \textrm{sinh}(u) \textrm{sin} (v) \textrm{cos} (\phi), \\
y &=& ce_{\textrm{p}}\, \textrm{sinh}(u)\textrm{sin}(v)\textrm{sin}(\phi), \\
z &=& ce_{\textrm{p}}\, \textrm{cosh}(u) \textrm{cos}(v),
\eea
where 
\beq
0 \le u < \infty, \quad 0 \le v \le \pi, \quad -\pi < \phi \le \pi.
\eeq
The set $(u,v,\phi)$ uniquely characterizes a point in the 3-dimensional space. It is straightforward to show that the metric coefficients are 
\beq\label{eq:p.metric}
h_u = h_v = ce_{\textrm{p}}\,\sqrt{\textrm{sinh}^2 u + \textrm{sin}^2 v}, 
\quad h_\phi = ce_{\textrm{p}}\, \textrm{sinh} u \, \textrm{sin} v
\eeq
using which the form for the Laplacian $\nabla^2 \Phi$ is readily obtained to be \cite{lebedev}
\bea\label{eq:laplacian}
\nabla^2\Phi &=& \frac{1}{c^2e_{\textrm{p}}^2(\textrm{sinh}^2u + \textrm{sin}^2v)}\times\nonumber\\
&&\left(
\frac{1}{\textrm{sinh}u}\frac{\delta}{\delta u}\left(\textrm{sinh}u \frac{\delta \Phi}{\delta u}\right)
+ \frac{1}{\textrm{sin}v}\frac{\delta}{\delta v}\left(\textrm{sin}v \frac{\delta \Phi}{\delta v}\right)\right)\nonumber\\
&& + \frac{1}{c^2e_{\textrm{p}}^2\textrm{sinh}^2u\,\textrm{sin}^2v}\frac{\delta^2\Phi}{\delta^2\phi}.
\eea

The prolate spheroidal shell in these coordinates is represented by the simple equation $u = u_0$, where $u_0$ is connected to the eccentricity via the relation 
\beq\label{eq:p.u0e}
\textrm{sech}u_0 = e_{\textrm{p}}.
\eeq
The region of space interior to the spheroid corresponds to the values $0\le u < u_0$ and the exterior region is represented by the $u > u_0$ domain. We begin by finding the electrostatic potential generated by the uniformly-charged prolate shell represented by the equation $u = u_0$. Since there is axial symmetry in the problem, the electrostatic potential will depend only on coordinates $u$ and $v$. Writing the solution as $\Phi(u,v) = U(u)V(v)$ and substituting it in the Laplace equation $\nabla^2\Phi = 0$, we find, upon using Eq.~\eqref{eq:laplacian}, that the variables separate and the functions $U$ and $V$ satisfy the differential equations:
\bea
&&\frac{1}{\textrm{sinh}u}\frac{\delta}{\delta u}\left(\textrm{sinh}u \frac{\delta U}{\delta u}\right) - nU = 0,\\
&&\frac{1}{\textrm{sin}v}\frac{\delta}{\delta v}\left(\textrm{sin}v \frac{\delta V}{\delta v}\right) + nV = 0.
\eea
A closer examination of these equations reveals the general solution for the potential to be:
\beq
\Phi(u,v) = \sum_{n=0}^{\infty}(A_n P_n(\textrm{cosh}u) + B_n Q_n(\textrm{cosh}u))P_n(\textrm{cos}v)
\eeq
where $n$ is an integer, $P_n$ and $Q_n$ are Legendre functions of the first and second kind respectively, and $A_n$ and $B_n$ are unknown coefficients. In order to ensure that the solutions are bounded in the interior and exterior regions of the spheroid, we find that $A_n$ must vanish in the domain $u > u_0$ and $B_n = 0$ in the region where $0 < u < u_0$. We thus have the following form for the potential inside and outside the oblate shell:
\bea
\Phi_{\textrm{in}}(u,v) &=& \sum_{n = 0}^{\infty} A_n P_n({\textrm{cosh}u}) P_n(\textrm{cos}v)\label{eq:p.in},\\
\Phi_{\textrm{out}}(u,v) &=& \sum_{n = 0}^{\infty} B_n Q_n({\textrm{cosh}u}) P_n(\textrm{cos}v)\label{eq:p.out}.
\eea

The potential must be continuous at the shell surface $u - u_0 = 0$, that is, $\Phi_{\textrm{in}}(u_0,v) = \Phi_{\textrm{out}}(u_0,v)$. This boundary condition leads to the relation 
\beq\label{eq:p.bc1}
A_n P_n({\textrm{cosh}u_0}) = B_n Q_n({\textrm{cosh}u_0})
\eeq
for $n = 0, 1, 2, \ldots$. Note that $\textrm{cosh}u_0 = 1/e_{\textrm{p}}$. The discontinuity in the normal component of the electric field at the charged surface provides another boundary condition:
\beq
-\hat{u}\cdot\nabla\Phi_{\textrm{out}} + \hat{u}\cdot\nabla\Phi_{\textrm{in}} = 4 \pi \sigma, \quad \textrm{at} \; u = u_0.
\eeq
Using the expression for the gradient in prolate spheroidal coordinates, the above equation becomes
\beq\label{eq:p.bc2}
\frac{1}{h_u}\frac{\delta\Phi_{\textrm{in}}}{\delta u}\Big\vert_{u=u_0}
-\frac{1}{h_u}\frac{\delta\Phi_{\textrm{out}}}{\delta u}\Big\vert_{u=u_0} = 4 \pi \sigma.
\eeq
Employing equations \eqref{eq:p.in}, \eqref{eq:p.out}, and \eqref{eq:p.metric}, the above boundary condition leads to the relation
\beq\label{eq:p.bc2n}
\begin{split}
\sum_{n = 0}^{\infty} &\left(A_n P'_n(1/e_{\textrm{p}})
- B_n Q'_n(1/e_{\textrm{p}})\right) P_n(\textrm{cos}v)  \\
&\quad\quad\quad\qquad\qquad= \frac{4 \pi \sigma c e_{\textrm{p}}}{\sqrt{1-e_{\textrm{p}}^2}}
\sqrt{1 - e_{\textrm{p}}^{2} \textrm{cos}^{2} v}.
\end{split}
\eeq

Eqs.~\eqref{eq:p.bc1} and \eqref{eq:p.bc2n} allow us to evaluate the undetermined functions $A_n$ and $B_n$. We first eliminate $B_n$ in favor of $A_n$ using Eq.~\eqref{eq:p.bc1} obtaining
\beq\label{eq:p.Bn}
B_n = \frac{A_n P_n(1/e_{\textrm{p}})}{Q_n(1/e_{\textrm{p}})}.
\eeq
Substituting $B_n$ from above in Eq.~\eqref{eq:p.bc2n}, we obtain 
\beq\label{eq:p.An}
\begin{split}
\sum_{n = 0}^{\infty} &\left(A_n P'_n(1/e_{\textrm{p}})
- \frac{A_n P_n(1/e_{\textrm{p}})}{Q_n(1/e_{\textrm{p}})} Q'_n(1/e_{\textrm{p}})\right) P_n(\textrm{cos}v)  \\
&\quad\quad\quad\qquad\qquad= \frac{4 \pi \sigma a e_{\textrm{p}}}{\sqrt{1-e_{\textrm{p}}^2}}
\sqrt{1 - e_{\textrm{p}}^{2} \textrm{cos}^{2} v}.
\end{split}
\eeq
Using the fact that the Wronskian of $P_n(z)$ and $Q_n(z)$ is given by 
\beq
W\left(P_n(z),Q_n(z)\right) = \frac{1}{1-z^2},
\eeq
Eq.~\eqref{eq:p.An} can be simplified to 
\beq
\sum_{n = 0}^{\infty} \frac{A_n P_n(\textrm{cos}v)}{Q_{n}(1/e_{\textrm{p}})} 
= \frac{4 \pi \sigma c \sqrt{1-e_{\textrm{p}}^2}}{e_{\textrm{p}}}\sqrt{1 - e_{\textrm{p}}^{2} \textrm{cos}^{2} v}.
\eeq
Multiplying both sides of the above equation with $P_l(\textrm{cos}v)\textrm{sin}v$ and using the orthogonality relation:
\beq
\int_{0}^{\pi} P_n(\textrm{cos}v) P_l(\textrm{cos}v) \textrm{sin}v \, dv = \frac{2}{2n+1}\delta_{nl}, 
\eeq
we obtain
\beq\label{eq:p.Anfinal}
A_n = \frac{2n+1}{2} 4\pi \sigma c \frac{\sqrt{1-e_{\textrm{p}}^2}}{e_{\textrm{p}}} \, Q_{n}(1/e_{\textrm{p}}) H_n(e_{\textrm{p}}),
\eeq
where $H_n(e_{\textrm{p}})$ is the integral 
\beq
H_{n}(e_{\textrm{p}}) = \int_{0}^{\pi} \sqrt{1 - e_{\textrm{p}}^{2} \textrm{cos}^{2} v} \,P_n(\textrm{cos}v) \textrm{sin}v \, dv.
\eeq
It is easily checked that for odd $n$ the integral in the above equation vanishes. Hence, we have $A_1 = A_3 = A_5 \ldots = 0$. Using $A_n$ in Eq.~\eqref{eq:p.Bn}, $B_n$ is known as well, and consequently from Eqs.~\eqref{eq:p.in} and \eqref{eq:p.out}, we obtain the desired electrostatic potential at any point in space. 

For the computation of the electrostatic energy, the knowledge of the potential on the shell surface suffices. Using Eqs.~\eqref{eq:p.in} and \eqref{eq:p.Anfinal}, we obtain the surface potential as
\beq\label{eq:p.phishell}
\begin{split}
\Phi_{\textrm{shell}}&(v,e_{\textrm{p}},c) = 
\frac{4\pi \sigma c \sqrt{1-e_{\textrm{p}}^2}}{e_{\textrm{p}}} \times \\
&\sum_{n\in \textrm{even}} \frac {2n+1}{2} \, P_{n}(1/e_{\textrm{p}}) Q_n(1/e_{\textrm{p}}) H_{n}(e_{\textrm{p}}) P_n(\textrm{cos}v),
\end{split}
\eeq
where the summation is over even integers $n = 0,2,4,\ldots$.
The electrostatic energy of a charged spheroidal shell can be written as
\beq\label{eq:Ushell}
U = \frac{1}{2}\int \sigma \, \Phi_{\textrm{shell}} \, dA. 
\eeq
The shell surface area element in the prolate spheroidal coordinates is given by 
\beq
dA = h_v h_{\phi} dv d\phi = c^2\sqrt{1-e_{\textrm{p}}^2}\sqrt{1 - e_{\textrm{p}}^2\textrm{cos}^2 v} \,\textrm{sin} v \, dv d\phi
\eeq
where the second equality follows from Eq.~\eqref{eq:p.metric}.
After substituting this expression for the area element in Eq.~\eqref{eq:Ushell}, using Eq.~\eqref{eq:p.phishell}, and changing the variable from $\sigma$ to $Q$, we obtain the expression in Eq.~\eqref{eq:p.U}. 

\section{Coulomb energy of uniformly-charged oblate spheroidal shells}\label{sec:oblate.energy}
We now derive the expression for the electrostatic energy of a uniformly-charged oblate spheroidal shell. As most of the steps involved in this derivation are analogous to the above derivation for the prolate case, we will present only the key steps of the procedure. We begin by employing the oblate spheroidal coordinates $u, v, \phi$, which are related to the Cartesian coordinates $x,y,z$ by
\bea
x &=& ae_{\textrm{o}}\, \textrm{cosh}(u) \textrm{sin}(v) \textrm{cos} (\phi), \\
y &=& ae_{\textrm{o}}\, \textrm{cosh}(u) \textrm{sin}(v) \textrm{sin}(\phi), \\
z &=& ae_{\textrm{o}}\, \textrm{sinh}(u) \textrm{cos}(v),
\eea
where 
\beq
0 \le u < \infty, \quad 0 \le v \le \pi, \quad -\pi < \phi \le \pi.
\eeq
The set $(u,v,\phi)$ uniquely characterizes a point in the 3-dimensional space. It is straightforward to show that the metric coefficients are 
\beq\label{eq:o.metric}
h_u = h_v = ae_{\textrm{o}}\,\sqrt{\textrm{sinh}^2 u + \textrm{cos}^2 v}, 
\quad h_\phi = ae_{\textrm{o}}\, \textrm{cosh} u \, \textrm{sin} v
\eeq
using which the form for the Laplacian $\nabla^2 \Phi = 0$ is readily obtained. The oblate spheroidal shell in these coordinates is represented by the simple equation $u = u_0$, where $u_0$ is connected to the eccentricity via the relation 
\beq\label{eq:o.u0e}
\textrm{sech}u_0 = e_{\textrm{o}}.
\eeq
The region of space interior to the spheroid corresponds to the values $0\le u < u_0$ and the exterior region is represented by the $u > u_0$ domain. 

Once again, we start by determining the electrostatic potential generated by this uniformly-charged oblate shell. Since this system has axial symmetry, the electrostatic potential created by the oblate spheroid will depend only on the coordinates $u$ and $v$. Employing separation of variables we can write the solution as $\Phi(u,v) = U(u)V(v)$, upon which the Laplace equation separates into two differential equations for $u$ and $v$. A closer examination of these equations reveals the general solution for the potential to be:
\beq
\Phi(u,v)= \sum_{n=0}^{\infty}(A_n P_n(i\,\textrm{sinh}u) + B_n Q_n(i\,\textrm{sinh}u))P_n(\textrm{cos}v),
\eeq
where $P_n$ and $Q_n$ are Legendre functions of the first and second kind respectively, and $A_n$ and $B_n$ are unknown coefficients. In order to ensure that the solutions are bounded in the interior and exterior regions of the spheroid, we find that $A_n = 0$ in the domain $u > u_0$, and $B_n = 0$ when $0 < u < u_0$. We thus have the following form for the potential inside and outside the oblate shell:
\bea
&&\Phi_{\textrm{in}} = \sum_{n = 0}^{\infty} A_n P_n({i\,\textrm{sinh}u}) P_n(\textrm{cos}v)\label{eq:o.in},\\
&&\Phi_{\textrm{out}} = \sum_{n = 0}^{\infty} B_n Q_n({i\,\textrm{sinh}u}) P_n(\textrm{cos}v)\label{eq:o.out}.
\eea

The boundary condition that the potential must be continuous at the shell surface $u - u_0 = 0$ leads to the relation 
\beq\label{eq:o.bc1}
A_n P_n({i\,\textrm{sinh}u_0}) = B_n Q_n({i\,\textrm{sinh}u_0})
\eeq
for $n = 0, 1, 2, \ldots$. Note that $\textrm{sinh}u_0 = (1/e_{\textrm{o}})\sqrt{1-e_{\textrm{o}}^2}$. The discontinuity in the normal component of the gradient of the electric potential at the charged surface provides another boundary condition which upon employing the expression for the gradient in oblate coordinates becomes:
\beq\label{eq:o.bc2n}
\begin{split}
\sum_{n = 0}^{\infty} &\left(A_n P'_n({i\,\textrm{sinh}u_0})
- B_n Q'_n({i\,\textrm{sinh}u_0})\right)i\textrm{cosh}u_0 P_n(\textrm{cos}v)  \\
&\quad\quad\quad\qquad\qquad= 4 \pi \sigma a \sqrt{1 - e_{\textrm{o}}^{2} \textrm{sin}^{2} v}.
\end{split}
\eeq
Using Eqs.~\eqref{eq:o.bc1} in \eqref{eq:o.bc2n}, we can eliminate $B_n$ in favor of $A_n$ and solve for the latter, obtaining
\beq\label{eq:o.An}
A_n = \frac{2n+1}{2} \frac{4\pi \sigma a\, i}{e_{\textrm{o}}} \, Q_{n}\left(i \, \frac{\sqrt{1-e_{\textrm{o}}^2}}{e_{\textrm{o}}}\right) I_{n}(e_{\textrm{o}}),
\eeq
where $I_n(e_{\textrm{o}})$ is the integral 
\beq
I_{n}(e_{\textrm{o}}) = \int_{0}^{\pi} \sqrt{1 - e_{\textrm{o}}^{2} \textrm{sin}^{2} v} \,P_n(\textrm{cos}v) \textrm{sin}v \, dv.
\eeq
As before, in arriving at this result we employed the property of the Wronskian of the Legendre polynomials and their orthogonality relation. It is easily checked that for odd $n$, $I_{n}(e_{\textrm{o}})$ vanishes, implying $A_1 = A_3 = A_5 \ldots = 0$. Using $A_n$ in Eq.~\eqref{eq:o.bc1}, $B_n$ can be evaluated as well and consequently from Eqs.~\eqref{eq:o.in} and \eqref{eq:o.out}, the electrostatic potential is known everywhere in space. 

For the computation of the electrostatic energy, the knowledge of the potential on the shell surface suffices. Using Eqs.~\eqref{eq:o.in} and \eqref{eq:o.An}, we obtain the surface potential as
\beq\label{eq:o.phishell}
\begin{split}
&\Phi_{\textrm{shell}}(v,e_{\textrm{o}},a) = 
\frac{4\pi \sigma a\, i}{e_{\textrm{o}}} \sum_{n \in \textrm{even}} \frac {2n+1}{2} \times \\
&P_{n}\left(i \, \frac{\sqrt{1-e_{\textrm{o}}^2}}{e_{\textrm{o}}}\right) Q_n\left(i\,\frac{\sqrt{1-e_{\textrm{o}}^2}}{e_{\textrm{o}}}\right) I_{n}(e_{\textrm{o}}) P_n(\textrm{cos}v).
\end{split}
\eeq

The shell surface area element in the oblate spheroidal coordinates is given by 
\beq\label{eq:areaelem.oblate}
dA = a^2 \sqrt{1 - e_{\textrm{o}}^2 \textrm{sin}^2 v} \, \textrm{sin} v \, dv d\phi.
\eeq
Using Eqs.~\eqref{eq:o.phishell} and \eqref{eq:areaelem.oblate} in Eq.~\eqref{eq:Ushell}, and changing the variable from $\sigma$ to $Q$, we obtain the energy expression in Eq.~\eqref{eq:o.U}.

\section{Coulomb energy of a uniformly-charged disc}\label{app:disc}
In this appendix, we derive the exact expression for the Coulomb energy of a uniformly-charged circular disc of radius $a$, total charge $Q$, and uniform charge density $\sigma = Q/(\pi a^2)$.  The potential on the surface of the disc as a function of $\rho$, the radial coordinate, has been derived in Ref.~\onlinecite{ciftja} and is given by:
\beq\label{eq:vrho}
V(\rho) = 4\sigma a E\left(\frac{\rho^2}{a^2}\right),
\eeq
where $E(m)$ is the complete elliptic integral of the second kind: 
\beq\label{eq:Em}
E(m) = \int_{0}^{\pi/2} \sqrt{1-m \,\textrm{sin}^2 \theta} \,d\theta.
\eeq
Note that $0\le\rho\le a$. This result can be used to obtain the electrostatic energy $U_{\textrm{disc}}$ of this disc. Interestingly, while the potential on the disc surface is only available as an elliptic integral, we will soon see that the total electrostatic energy of the disc reduces to a simple form. 

Starting with the definition of the electrostatic energy, $U = (1/2)\int\sigma V dA$, we have
\beq
U_{\textrm{disc}} = \frac{1}{2}\sigma\int_{0}^{a} V(\rho) 2\pi \rho d\rho.
\eeq
Substituting $V(\rho)$ from Eq.~\eqref{eq:vrho} and using Eq.~\eqref{eq:Em}, we obtain
\beq
U_{\textrm{disc}} = 4\pi\sigma^2 a \int_{0}^{a} d\rho \int_{0}^{\pi/2} d\theta \sqrt{1- \frac{\rho^2}{a^2} \,\textrm{sin}^2 \theta} \, \rho.
\eeq
Carrying out the integral with respect to $\rho$ first by employing the substitution $t = 1 - \rho^2 \textrm{sin}^2\theta/a^2	$, we obtain
\beq
U_{\textrm{disc}} = 4\pi\sigma^2 \frac{a^3}{3} \int_{0}^{\pi/2} \frac{1-\textrm{cos}^3\theta}{\textrm{sin}^2\theta}\, d\theta.
\eeq
The integral over $\theta$ equates to 2 leading to the following expression for the energy:
\beq
U_{\textrm{disc}} = \frac{8\pi}{3}\sigma^2 a^3.
\eeq
To our best knowledge, this expression is reported here for the first time.

%

\end{document}